\documentclass{iopart}
\usepackage{iopams}
\usepackage{graphicx}

\newcommand{\MDP}{\hbox{MDP--$p$\ }} %
\newcommand{\MDPns}{\hbox{MDP--$p$}}%
\newcommand{\dilog}{\mathrm{Li_2}}%

\begin{document}

\title[Limit shapes and the largest part in the minimal difference
partitions]{Integer partitions and exclusion statistics: Limit shapes and
  the largest part of Young diagrams}

\author{Alain Comtet$^{1,2}$, Satya N. Majumdar$^{1}$, St\'ephane
  Ouvry$^{1}$ and Sanjib Sabhapandit$^{1}$}

\address{$^{1}$Laboratoire de Physique Th\'eorique et Mod\`eles
  Statistiques,\\
  Universit\'e de Paris-Sud, CNRS UMR 8626, 91405 Orsay
  Cedex, France \\
  $^{2}$ Institut Henri Poincar\'e, 11 rue Pierre et Marie Curie, 75005
  Paris, France}

\begin{abstract}
  We compute the limit shapes of the Young diagrams of the minimal
  difference $p$ partitions and provide a simple physical interpretation
  for the limit shapes. We also calculate the asymptotic distribution of
  the largest part of the Young diagram and show that the scaled
  distribution has a Gumbel form for all $p$.  This Gumbel statistics for
  the largest part remains unchanged even for general partitions of the
  form $E=\sum_i n_i i^{1/\nu}$ with $\nu>0$ where $n_i$ is the number of
  times the part $i$ appears.
\end{abstract}

\noindent Journal-ref: {\it J. Stat. Mech. (2007) P10001}

\noindent\rule{\hsize}{2pt}
\tableofcontents
\noindent\rule{\hsize}{2pt}

\title[Limit shapes and the largest part in the minimal difference
partitions]\maketitle

\section{Introduction}

Exclusion statistics~\cite{Haldane, DO1, DO2, Wu, MS1, MS2, Isakov,
  Bergere}---a generalization of Bose and Fermi statistics---can be defined
in the following thermodynamical sense. Let $Z(\beta, z)$ denote the grand
partition function of a quantum gas of particles at inverse temperature
$\beta$ and fugacity $z$. Such a gas is said to obey exclusion statistics
with parameter $0\le p\le 1$, if $Z(\beta, z)$ can be expressed as an
integral representation
\begin{equation}
  \ln Z(\beta,z)=\int_0^{\infty} {\tilde \rho}(\epsilon)
  \ln y_p\left(z\rme^{-\beta\epsilon}\right)\,
  \rmd\epsilon,
\label{thermo}
\end{equation}
where ${\tilde \rho}(\epsilon)$ denotes a single particle density of
states and the function $y_p(x)$, which encodes fractional statistics, is
given by the solution of the  equation
\begin{equation}
y_p(x)- x\, y_p^{1-p}(x)=1.
\label{func1}
\end{equation}
In the cases $p=0$ and $p=1$, substituting $y_p(x)$ explicitly in
\eref{thermo} yield the standard grand partition functions of
non-interacting bosons and fermions respectively.  The fractional exclusion
statistics with parameter $0< p< 1$ (that corresponds to an interacting
gas) smoothly interpolates between these two extreme cases.  Two known
microscopic quantum mechanical realizations of exclusion statistics are the
Lowest Landau Level (LLL) anyon model \cite{DO1,DO2} and the Calogero model
\cite{MS2,Isakov}, with ${\tilde \rho}(\epsilon)$ being, respectively, the
LLL density of states and the free one-dimensional density of states.

\begin{figure}
\centering
\includegraphics[height=6cm]{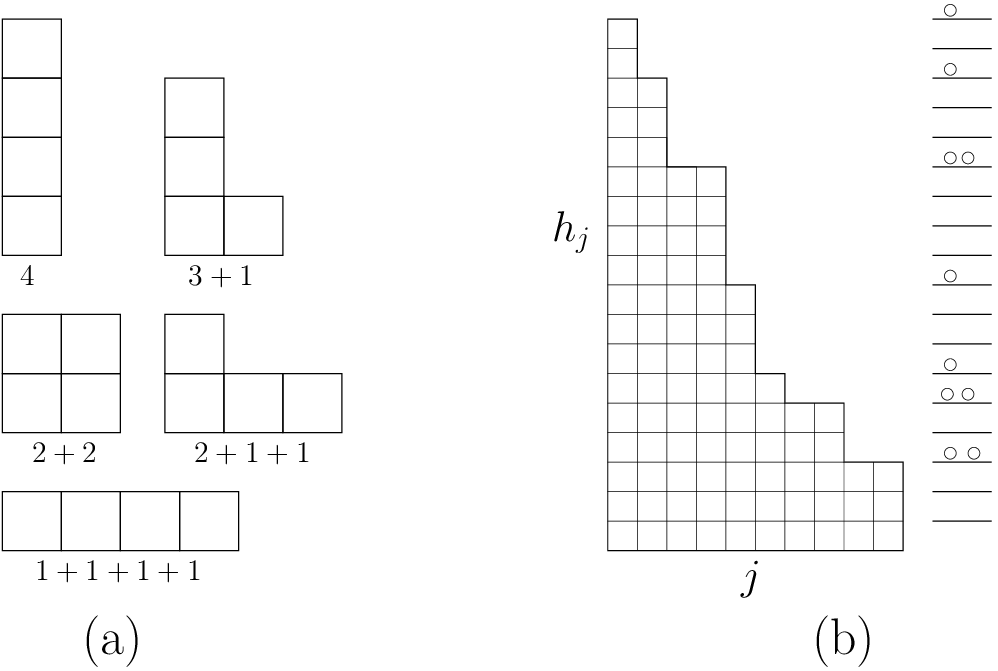}
\caption{\label{youngD} (a) All the Young diagrams for the partitions of 4.
  (b) The Young diagram of the partition $91=18+16+13+13+9+6+5+5+3+3$, and
  the corresponding configuration of non-interacting bosons occupying
  energy levels $\epsilon_i=i$ for $i=1,2,\ldots,18$.}
\end{figure}

It is well known that a gas of non-interacting bosons ($p=0$) or fermions
($p=1$) occupying a single particle equidistant spectrum both have a
combinatorial interpretation in terms of the integer partition
problem~\cite{Andrews}. A partition of a positive integer $E$ is a
decomposition of $E$ as a sum of a nonincreasing sequence of positive
integers $\{h_j\}$, i.e., $E=\sum_j h_j$ such that $h_j \ge h_{j+1}$, for
$j=1,2\ldots$.  For example, $4$ can be partitioned in $5$ ways: $4$,
$3+1$, $2+2$, $2+1+1$, and $1+1+1+1$.  Partitions can be graphically
represented by Young diagrams (also called Ferrers diagrams), where $h_j$
corresponds to the height of the $j$-th column (see \fref{youngD}).
In the Young diagram of a given partition of $E$, if $n_i$ denotes the
number of columns having heights equal to $i$, then clearly $E=\sum_i
n_i\epsilon_i$ ---which can now be interpreted as the total energy of a
non-interacting quantum gas of bosons where $\epsilon_i=i$ for
$i=1,2,\ldots,\infty$ represent equidistant single particle energy levels
and $n_i=0,1,2,\ldots,\infty$ represents the occupation number of the
$i$-th level (see \fref{youngD}(b)). On the other hand, if one expresses a
positive integer $E$ as a sum of strictly decreasing sequence of positive
integers, i.e. $E=\sum_j h_j$ such that $h_j > h_{j+1}$ (e.g. allowed
partitions of 4 are: $4$ and $3+1$), then the restricted partition problem
corresponds to a non-interacting quantum gas of fermions, for which
$n_i=0,1$. In the partitioning problems if one restricts the number of
summands to be $N$, then clearly $N=\sum_i n_i$ represents the total number
of particles. For example, if $E=4$ and $N=2$, the allowed partitions are
$3+1$ and $2+2$ in the unrestricted problem, whereas the only allowed
restricted partition is $3+1$.  The number $\rho(E,N)$ of ways of
partitioning $E$ into $N$ parts is simply the micro-canonical partition
function of a gas of quantum particles with total energy $E$ and total
number of particles $N$:
\begin{equation}
  \rho(E,N) =\sum_{\{n_i\}} \delta\left(E-\sum_{i=1}^{\infty} n_i
    \epsilon_i\right)\,
  \delta\left(N-\sum_{i=1}^{\infty} n_i\right).
\label{ren1}
\end{equation}
The grand partition functions, i.e., $Z(\beta,z)=\sum_N\sum_E z^N
\rme^{-\beta E}\rho(E,N)$, for the unrestricted and restricted partitions
are $Z(\beta,z)=\prod_{i=1}^{\infty} (1-z\rme^{-\beta i})^{-1}$ and
$Z(\beta,z)=\prod_{i=1}^{\infty} (1+z\rme^{-\beta i})$ and hence $\ln
Z(\beta,z)$ in the limit $\beta\rightarrow 0$ and
$\tilde{\rho}(\epsilon)=1$ reduce to \eref{thermo} with $p=0$ and $p=1$
respectively.

Unlike Bose and Fermi statistics which describes non-interacting particles,
for a quantum gas obeying exclusion statistics with parameter $0<p<1$, it
is a priori not obvious how to provide a combinatorial description, since
the underlying physical models with exclusion statistics describe
interacting systems.  However it has recently been shown~\cite{CMO} that a
combinatorial description of exclusion statistics is possible in terms of a
generalized partition problem known as the minimal difference $p$ partition
(\MDPns), which we will define in the next section.  Even though the
parameter $p$ in \MDP is an integer, in \cite{CMO} it has been shown that,
when one analytically continues the results to non-integer values of $p$,
for $0<p<1$, and in the limit $\beta\to 0$, the \MDP corresponds to a gas
of quantum particles obeying exclusion statistics.  This correspondence
between exclusion statistics and \MDP motivates us to investigate some
other aspects of the \MDP problem in this paper.

\section{Problems and outline}
\label{outline}

\begin{figure}
\centering
\includegraphics[height=6cm]{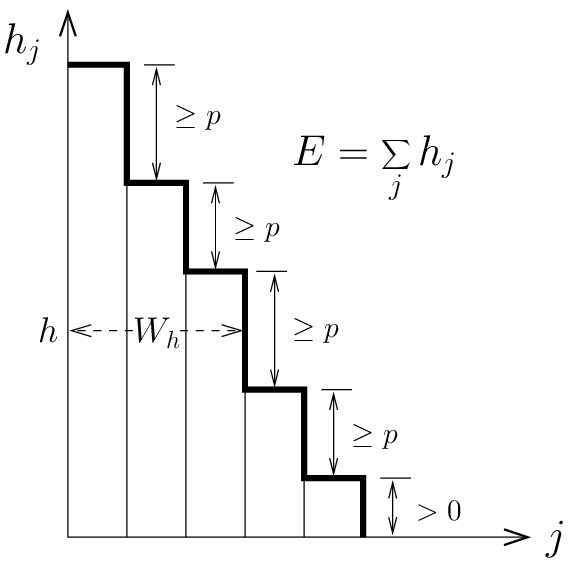}
\caption{\label{mdp} A typical Young diagram for \MDP problem. The thick
  solid border shows the height profile. $W_h$ is the width of the Young
  diagram at a height $h$, i.e., $W_h$ is the number of columns whose
  heights $\ge h$.}
\end{figure}

In the \MDP problem, a positive integer $E$ is expressed as a sum of
positive integers $E=\sum_j h_j$ such that $h_j-h_{j+1}\ge p$ (see
\fref{mdp}).  Therefore, $p=0$ corresponds to unrestricted partitions and $
p=1$ to restricted partitions into distinct parts.  The shortest part in
the \MDP problem is usually taken to be $\ge1$.  However, for the
calculation of certain specific quantities in this model, it is useful to
consider a somewhat generalized version with the shortest part $\ge s$,
where $s$ is considered to be a variable. The grand partition function of
this problem was obtained recently in~\cite{CMO}, which is given by
\eref{thermo} with constant density of states ${\tilde \rho}(\epsilon)=1$
and the lower limit of integration being $s$.

One may also think of the \MDP in terms of a quantum system consisting of
equidistant energy levels $\epsilon_i=i$ for $i=1,2,\ldots,\infty$.  Now a
given height $h_j=i$ corresponds the energy level $\epsilon_i=i$ and the
number of columns with height $i$ is the occupation number $n_i$.  Since
the difference between two consecutive heights in the \MDP must be at least
$p$, the gap between two adjacent occupied energy levels must be at least
$p$.  Clearly for $p=0$ this gap is zero, and hence each level can be
occupied by any number of particles (bosons).  For $p=1$, each level can be
occupied by at most one particle (fermions).  Again for $p>1$ a level can
be occupied by at most one particle.  However, in this case, when a energy
level is occupied by a particle, the adjacent $p-1$ levels must remain
unoccupied.

One major issue in the partition problem is to study the limit shape, i.e.,
the average height profile of an ensemble of Young diagrams with a fixed
but large $E$. The shape (height profile) can be defined by the width $W_h$
of the Young diagram at a height $h$ (see \fref{mdp}). In other words,
$W_h$ is the number of columns of the Young diagram whose height is greater
than or equal to $h$.  In this corresponding quantum system, $W_h$
represents the total number of particles occupying energy levels above $h$.

The height profile of the Young diagram of the unrestricted partition
($p=0$) was first studied by Temperley, who was interested in determining
the equilibrium profile of a simple cubic crystal grown from the corner of
three walls at right angles.  The two dimensional version of the problem
---where walls (two) are along the horizontal and the vertical axes and $E$
``bricks'' (molecules) are packed into the first quadrant one by one such
that each brick, when it is added, makes two contact along faces---
corresponds to the $p=0$ partition problem.  Temperley~\cite{Temperley}
computed the equilibrium profile of this two dimensional crystal.  More
recently the investigation of the limit shape of random partitions has been
developed extensively by Vershik~\cite{Vershik,Vershik2,Vershik3} and
collaborators.  The case of uniform random partitions was treated by
Vershik who proved for the bosonic ($p=0$) as well as the fermionic ($p=1$)
case that the rescaled $h/\sqrt{E}$ vs.  $W_h/\sqrt{E}$ curves converge to
limiting curves when $E\rightarrow\infty$, and obtained these limit shapes
explicitly.  These results were extended by Romik~\cite{Romik} to the \MDP
for $p=2$. In this paper we compute the following two quantities:

\begin{enumerate}
\item The limit shape of the Young diagrams of the \MDP for any $p$, from
  which the previously obtained results for $p=0,1,2$ follow as special
  cases.

\item The distribution of the largest part of the Young diagrams of the
  \MDP problem for all $p$, whereas the earlier result existed only for the
  $p=0$ case~\cite{Erdos}.
\end{enumerate}

The average height profile $\langle W_h\rangle$ of the Young diagrams of
the partitions of a given integer $E$ is easier to compute in the grand
canonical ensemble.  Therefore one requires a restricted grand partition
function $Z_h(\beta,z)$ which counts the columns whose heights $\ge h$, and
the full grand partition function $Z(\beta,z)$ which counts all the
columns. From the restricted grand partition function one finds $\langle
W_h\rangle=z\frac{\partial}{\partial z} \ln Z_h(\beta,z)|_{z=1}$.  For
given large $E$, the parameter $\beta$ is fixed by the relation
$E=-\frac{\partial}{\partial \beta} \ln Z(\beta,1)$.

On the other hand, to compute the number of partitions $\rho_p(E,l)$ of an
integer $E$ such that the largest part $\le l$, it is useful to consider
the partition function $Z_l(\beta)=\sum_E \rme^{-\beta E} \rho_p(E,l)$
first. Formally $\rho_p(E,l)$ can be obtained by inverting $Z_l(\beta)$
with respect to $\beta$, and for large $E$ the asymptotic behavior of
$\rho_p(E,l)$ is obtained from the saddle point approximation, where the
parameter $\beta$ is fixed in terms of given $E$ by the saddle point
relation $E=-\frac{\partial}{\partial \beta} \ln Z_l(\beta)$.

Thus, it is useful to consider a more general restricted grand partition
function $Z(\beta,z,l,s)$ that counts the columns whose heights lie between
$s$ and $l$.  All the other partition functions we need for our
calculations can be obtained from $Z(\beta,z,l,s)$ by taking various limits
on $s$ and $l$.  For example, by putting $s=1$ and taking the limit
$l\rightarrow\infty$ one obtains $Z(\beta,z)$.  Similarly $s=h$ and the
limit $l\rightarrow\infty$ gives $Z_h(\beta,z)$ and putting $s=1$ and $z=1$
gives $Z_l(\beta)$.  As we will see later in \eref{beta} and \eref{saddle1}
that $\beta\sim E^{-1/2}$ for large $E$. Therefore, hereafter we will work
in the limit $\beta\rightarrow 0$.

The rest of the paper is organized as follows.  We first obtain the
generalized grand partition function $Z(\beta,z,l,s)$ of the \MDP problem
in the next section.  In \sref{limit shape} we compute the limit shapes of
the Young diagrams and also provide a simple physical interpretation of the
result.  In \sref{largest part} we calculate the distribution of the
largest part of the \MDP.  Finally, we conclude with a summary and some
remarks in \sref{summary}.

\section{Restricted grand partition function of \MDP problem}
\label{GPF}

Let $\rho_p(E,N,l,s)$ be the number of ways of partitioning an integer $E$
into $N$ parts in the \MDP problem such that the largest part is at most
$l$ and the smallest part is at least $s$, i.e., $E=\sum_{j=1}^{N} h_j$
such that $h_1\le l$, $h_{j+1}\le h_j-p$ for all $j=1,2,\ldots,N-1$, and
$h_N\ge s$.  Then clearly, $[\rho_p(E,N,l,s) - \rho_p(E,N,l-1,s)]$ gives
the number of \MDP of $E$, such that the largest part is exactly equal to
$l$, and smallest part is at least $s$.  Now, by eliminating the first part
$h_1=l$ from the partition one immediately realizes that the above number
is precisely $\rho_p(E-l,N-1,l-p,s)$, i.e., the number of \MDP of $E-l$
into $N-1$ parts such that the largest part is at most $l-p$ and the
smallest part is at least $s$.  Therefore, one has the recursion relation
\begin{equation}
  \rho_p(E,N,l,s) = \rho_p(E,N,l-1,s) + \rho_p(E-l,N-1,l-p,s).
  \label{recursion1}
\end{equation}
Following similar reasoning one can also derive another recursion relation
in terms of the smallest part $s$,
\begin{equation}
  \rho_p(E,N,l,s) = \rho_p(E,N,l,s+1) + \rho_p(E-s,N-1,l,s+p).
  \label{recursion2}
\end{equation}
It follows from \eref{recursion1} and \eref{recursion2} that the grand
partition function $Z(\beta,z,l,s)=\sum_N\sum_{E} z^N \rme^{-\beta
  E}\rho_p(E,N,l,s)$ satisfies the  recursion relations:
\begin{eqnarray}
\label{frel1}
  Z(\beta,z,l,s)=Z(\beta,z,l-1,s)+z \rme^{-\beta l} Z(\beta,z,l-p,s),\\
\label{frel2}
  Z(\beta,z,l,s)=Z(\beta,z,l,s+1)+z \rme^{-\beta s} Z(\beta,z,l,s+p).
\end{eqnarray}
From these equations, it is evident that in the scaling limit
$\beta\rightarrow 0$, and both $s$ and $l$ large, the correct scaling
variables are $\beta s$ and $\beta l$, so that $\beta s$ and $\beta l$
remain finite.  One knows from the statistical mechanics that the free
energy $\beta^{-1} \ln Z(\beta,z,l,s)$ becomes a function of the only the
scaling variables in the limit $\beta\rightarrow 0$. Therefore in this
limit it is natural to expect
\begin{equation}
  Z(\beta,z,l,s) \approx
  \exp\left(\frac{1}{\beta} \Phi(\beta l,\beta s,z) \right).
\label{ansatz}
\end{equation}
Now to determine the scaling function $\phi(\beta l,\beta s,z)$, we
substitute the ansatz \eref{ansatz} in \eref{frel1} and \eref{frel2}, and
then expand $\Phi(\beta l-\beta,\beta s,z)$ and $\Phi(\beta l-\beta p,\beta
s,z)$ about $\beta l$, and $\Phi(\beta l,\beta s+\beta,z)$ and $\Phi(\beta
l,\beta s +\beta p,z)$ about $\beta s$, respectively in Taylor series up to
first order, which yields the equations:
\begin{eqnarray}
\label{diffeq1}
\fl\qquad
\exp(-\Phi_{\beta l}) &+ z \rme^{-\beta l} \exp(-p\Phi_{\beta l})&=1,
\quad\mbox{where}\quad 
\Phi_{\beta l}=\frac{\partial}{\partial u} \Phi(u,\beta s,z)
\Big|_{u=\beta l},\\
\label{diffeq2}
\fl\qquad
\exp(\Phi_{\beta s}) &+ z \rme^{-\beta s} \exp(p\Phi_{\beta s})&=1, 
\quad\mbox{where}\quad 
\Phi_{\beta s}=\frac{\partial}{\partial v} \Phi(\beta l,v,z)
\Big|_{v=\beta s}.
\end{eqnarray}
It is evident from \eref{diffeq1} and \eref{diffeq2}, that $\Phi_{\beta l}$
and $\Phi_{\beta s}$ are function of the arguments $z \rme^{-\beta l}$ and
$z \rme^{-\beta s}$ respectively, and the solutions are
\begin{equation}
  \Phi_{\beta l}=\ln\,y_p\left(z\rme^{-\beta l}\right)
\quad\mbox{and}\quad
  \Phi_{\beta s}=-\ln\,y_p\left(z\rme^{-\beta s}\right)
\label{substitution}
\end{equation}
where $y_p(x)$ satisfies the equation $y_p(x)- x\, y_p^{1-p}(x)=1$, which
is the same equation \eref{func1} one encounters in exclusion statistics.
\Eref{substitution} implies,
\begin{math}
  \Phi(u,v,z)=\int_{v}^{u} \ln
  y_p\left(z\rme^{-\epsilon}\right)\, \rmd\epsilon.
\end{math}
Therefore, \eref{ansatz} yields
\begin{equation}
  \ln Z(\beta,z,l,s) =\frac{1}{\beta} \int_{\beta s}^{\beta l} 
\ln y_p\left(z\rme^{-\epsilon}\right)\,  \rmd\epsilon, 
\label{general gp}
\end{equation}
i.e. \eref{thermo} with constant density of states ${\tilde
  \rho}(\epsilon)=1$, and the lower and upper limits of integration being
$s$ and $l$ respectively.  This is the key equation, using which we compute
the limit shapes and the largest parts of the Young diagrams in \sref{limit
  shape} and \sref{largest part} respectively.  The limit $\beta
l\rightarrow \infty$ also provides a simpler derivation of an earlier
result~\cite{CMO}, which showed a link between the exclusion statistics and
the \MDP problem.

\section{Limit shapes of Young diagrams}
\label{limit shape}

Let us consider all the \MDP of an integer $E$ with uniform measure. Then
the number of columns having height between $s$ and $l$, averaged over all
the Young diagrams of the \MDP of $E$, is obtained from \eref{general gp}
as
\begin{equation}
  \fl\qquad
  \left\langle N_s^l (z)\right\rangle 
  = z\frac{\partial}{\partial z} \ln Z(\beta,z,l,s)
  =\frac{1}{\beta} \left[\ln y_p\left(z\rme^{-\beta s}\right)
    -\ln y_p\left(ze^{-\beta l}\right)\right].
  \label{number of columns}
\end{equation}
Now to obtain the parameter $\beta$ in terms of the given large integer $E$
one again uses \eref{general gp} with the limits $\beta
l\rightarrow\infty$, $\beta s\rightarrow 0$, and $z=1$, i.e.,
\begin{equation}
  \fl\qquad
  E=-\frac{\partial}{\partial \beta} \ln Z(\beta,1,\infty,0)
  =\frac{b^2(p)}{\beta^2},
  \quad\mbox{where}\quad 
  b^2(p) = \int_{0}^{\infty} 
  \ln y_p\left(\rme^{-\epsilon}\right)\,  \rmd\epsilon
\label{beta}
\end{equation}
is a constant which depends on the parameter $p$.

The average shape or the height profile of the Young diagrams $\langle
W_h\rangle$ is simply given by \eref{number of columns} with $s=h$,
$l\rightarrow\infty$ and $z=1$, i.e.,
\begin{equation}
  \beta \left\langle W_h\right\rangle = \ln y_p\left(\rme^{-\beta
      h}\right), 
  \quad\mbox{where}\quad 
\beta=\frac{b(p)}{\sqrt{E}}.
\label{shape}
\end{equation}
For instance for $p=0,1$ and $2$, solving \eref{func1} yields
$y_0(x)=1/(1-x)$, $y_1(x)=(1+x)$, and $y_2(x)=\left[1+\sqrt{1+4x}\right]/2$
respectively. From which using \eref{beta} one finds $b(0)=\pi/\sqrt{6}$,
$b(1)=\pi/\sqrt{12}$ and $b(2)=\pi/\sqrt{15}$ in agreement with the earlier
known results \cite {Vershik, Romik}.

\begin{figure}
  \centering
\includegraphics[height=6cm]{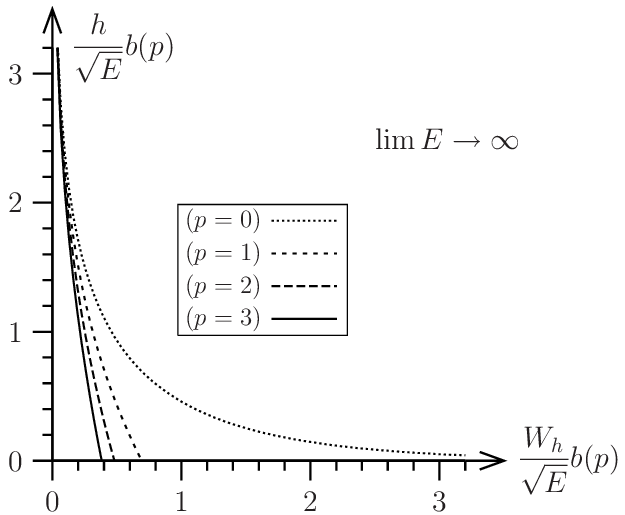}
\caption{\label{shapefig} Limit shapes for the minimal difference $p$
  partitions with $p=0,1,2$, and $3$, where $b(0)=\pi/\sqrt{6}$,
  $b(1)=\pi/\sqrt{12}$, $b(2)=\pi/\sqrt{15}$, and $b(3)=0.752617\ldots$.}
\end{figure}

The fluctuation about the average shape can be computed from \eref{general
  gp} using
\begin{equation}
  \langle W_h^2\rangle - \langle
  W_h\rangle^2
  = z\frac{\partial}{\partial z}z\frac{\partial}{\partial z} 
  \ln Z(\beta,z,\infty,h)\Big|_{z=1},
\end{equation}
which gives
\begin{equation}
  \beta^2\left[  \langle W_h^2\rangle - \langle
    W_h\rangle^2\right]
  =\beta \rme^{-\beta h} 
  \left[\frac{y_p'\left(\rme^{-\beta h}\right)}{
    y_p\left(\rme^{-\beta h}\right)} \right],
\end{equation}
where $y_p'(x)$ denotes the derivative of $y_p(x)$ with respect to its
argument. This formula shows that the random variable $\beta W_h$ is
strongly peaked around its mean value. Therefore, the curve $W_h/\sqrt E$
as a function of $h/\sqrt E$ converges to a limit curve when
$E\rightarrow\infty$ (strictly speaking, to prove the existence of a limit
curve, one needs to show that all the moments around the mean vanish when
$E\rightarrow \infty$, which Vershik~\cite {Vershik2} showed for $p=0$ and
$p=1$).  Therefore hereafter we may replace $ \langle \beta W_h \rangle$ by
$\beta W_h$.

Using \eref{func1} and \eref{shape}, one can express $h$ in terms of $ W_h$
as,
\begin{equation}
  h =
  -\frac{1}{\beta}\ln \left(1-\rme^{-\beta W_h } \right) 
  -p  W_h.
\label{reverse shape}
\end{equation}
Introducing the scaling variables $x=W_h/\sqrt E$ and $y=h/\sqrt E$, using
\eref{beta} and taking $E\rightarrow\infty$, yields the equation of the
limit shape
\begin{equation}
  y=-\frac{1}{b(p)} \ln (1-\e^{-xb(p)}) -px.
\end{equation}
\Fref{shapefig} shows the limit shapes for the \MDP with $p=0,1,2$, and
$3$.

\begin{figure}
  \centering
  \includegraphics[width=\hsize]{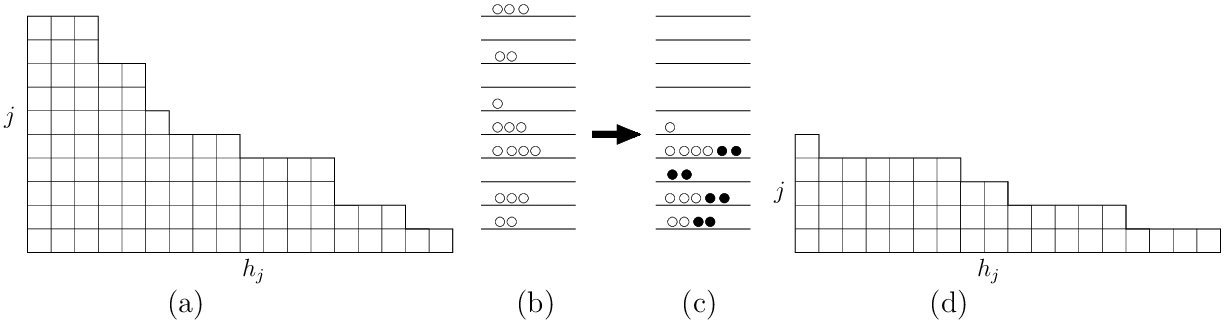}
  \caption{\label{transposeYD} (a) Transposed Young diagram for the
    unrestricted partition $91=18+16+13+13+9+6+5+5+3+3$. (b) Represents (a)
    in terms of non-interacting bosons (represented by \opencircle)
    occupying energy levels $\epsilon_i=i$ for $i=1,2,\ldots,10$. (c) The
    configuration obtained from the bosonic configuration (b) by
    transferring particles from the higher levels to the lower levels such
    that in the final configuration all the levels below the highest
    occupied level $\epsilon_5=5$ receive 2 new particles (represented by
    \fullcircle) each, where \opencircle represents the particles
    originally present in the initial bosonic configuration.  (d) The Young
    diagram corresponding to the configuration (c). This is the transposed
    Young diagram of the partition $49=18+14+9+7+1$, in the minimal
    difference 2 partition problem.}
\end{figure}

\Eref {reverse shape} has a simple physical interpretation which we explain
below.  For $ p=0$, any transposed Young diagram (see \fref{transposeYD})
provides a valid unrestricted partition. Therefore the transposed diagram
also corresponds to a non interacting system of bosons occupying single
particle equidistant energy levels. However this is no more true when
$p>0$. In this case the transposed Young diagram(see \fref{transposeYD})
corresponds to a quantum system where there is a certain energy level
(which differs from one realization to another) which is occupied by at
least one particle, and above which all the levels are empty, and below
which each of the levels must be occupied by at least $p$ particles.
Therefore, $h$ in the limit shape expression \eref{reverse shape}
represents the number of particles above the energy level $W_h $. For
bosons with total energy $E$, this number is precisely given by
\eref{reverse shape} with $p=0$ and $\beta$ has to be determined in terms
of $E$.  Now, a configuration for $p>0$, can be obtained from a bosonic
configuration by transferring particles from the higher energy levels to
the lower ones such that, in the final configuration, levels below the
highest occupied level (which has at least one particle) receive exactly
$p$ new particles each.  Clearly, in the final configuration obtained by
this procedure, each of the levels below the highest occupied level has at
least $p$ particles.  However, since transferring a particle from a higher
energy level to a lower one decreases energy of the system, to obtain a
configuration for $p>0$ with energy $E$ requires the initial bosonic
configuration to be at a higher energy (i.e., lower inverse temperature
$\beta$) than $E$.  Now, while going from a initial bosonic configuration
to a configuration for $p>0$, one transfers total of $p W_h$ particles from
the levels above $ W_h$ to below (i.e., $p$ particles to each level), the
average number number of particles above level $ W_h$ decreases from the
corresponding bosonic system ($p=0$) precisely by $p W_h$, which is exactly
the content of \eref{reverse shape} .  In fact, $\beta$ in \eref{reverse
  shape} can directly be determined by using condition $h\ge 0$ and the
normalization $\int_0^{ W_h^*} h\left( W_h \right)\, \rmd W_h =E$, where $
W_h^*$ is the solution of the equation $h\left( W_h^*\right)=0$. Writing
$\exp\left(\beta W_h^*\right) = y^*$, it satisfies $y^* -y^{* 1-p}=1$, and
in terms of $y^*$ one finds
\begin{equation}
  \beta =\frac{b(p)}{\sqrt{E}}
  \quad\mbox{with}\quad
  b^2(p)= \frac{\pi^2}{6} -\dilog (1/y^*)-\frac{p}{2} (\ln y^*)^2,
\label{b(p)}
\end{equation}
where $\dilog(z)=\sum_{k=1}^\infty z^k k^{-2}$ is the dilogarithm function.
The expression for $b(p)$ in \eref{b(p)} also follows directly from its
integral representation given in \eref{reverse shape}.

\section{Largest part of Young diagrams}
\label{largest part}

\Eref{general gp} also allows one to compute the distribution of the
largest part (i.e., the largest height in the Young diagram) in the \MDP
problem. Let $\rho_p(E,l)$ be the number of partitions of the integer $E$
in \MDP problem, such that the largest part is at most $l$. Clearly,
$\rho_p(E) =\rho_p(E,l\rightarrow\infty)$ gives the total number of
partitions of $E$ and since the partitions are distributed with a uniform
measure $C_p(l|E)=\rho_p(E,l)/\rho_p(E)$ gives the cumulative distribution
of the largest height $l$. Note that the partition function
$Z_l(\beta)=\sum_{E} \rme^{-\beta E}\rho_p(E,l)$ in the limit
$\beta\rightarrow 0$ is obtained from \eref{general gp} by simply taking
the limit $\beta s\rightarrow 0$ and $z=1$. Therefore, formally inverting
the Laplace transform (in the limit $\beta\rightarrow 0$, the sum over $E$
in the partition function of $Z_l(\beta)$ can be replaced by an integral),
one can write
\begin{equation}
  \rho_p(E,l) = \frac{1}{2\pi i} \int_{\gamma-i\infty}^{\gamma+i\infty}  
  \exp\left[S_{E,l}(\beta) 
  \right]\,\rmd \beta,
\end{equation}
where $\gamma$ is a real constant chosen such that all singularities of
integrand are to the left of the vertical contour in the complex plane, and
the action
\begin{equation}
  S_{E,l}(\beta)=\beta E+
  \frac{1}{\beta}\int_{0}^{\beta l} \ln y_p\left(\rme^{-\epsilon}\right)\,
  \rmd\epsilon.
  \label{action}
\end{equation}
For large $E$, the leading asymptotic behavior of $\rho_p(E,l)$ can be
obtained from the saddle point approximation. Maximizing the action with
respect to $\beta$, i.e., setting $\partial S/\partial \beta =0$ gives the
saddle point equation
\begin{equation}
  \beta^2 E=
  \int_{0}^{\beta l} 
  \ln y_p\left(\rme^{-\epsilon}\right)\,
  \rmd\epsilon -\beta l \ln y_p\left(\rme^{-\beta l}\right).
  \label{saddle1}
\end{equation}
For large $E$, the saddle point $\beta^*$ is obtained implicitly solving
the above equation and by substituting it back in the action
$S_{E,l}(\beta^*)$. Thus, to leading order,
\begin{equation}
  \rho_p(E,l) \approx \exp\left[S_{E,l}(\beta^*)\right],
\label{saddle2}
\end{equation}
where $S_{E,l}(\beta^*)$ can be written as
\begin{equation}
  S_{E,l}(\beta^*)\approx
  \frac{1}{\beta^*}\left[2 \int_0^{\beta^* l} 
    \ln y_p\left(\rme^{-\epsilon}\right)\,
    \rmd\epsilon -\beta^* l \ln y_p\left(\rme^{-\beta^* l}\right)\right].
  \label{entropy1}
\end{equation}

It is evident from the above equations that,  in terms of $l$ and
$E$, one has the scaling form $S_{E,l}(\beta^*)=\sqrt{E}\,
g_p(l/\sqrt{E})$, where the scaling function $g_p(x)$ can be determined
as follows.  We set $l/\sqrt{E}=x$ and $\beta^* l= H_p(x)$. In terms of
these scaling variables, from the saddle point solution of \eref{saddle1}
and the entropy \eref{entropy1} one has
\begin{eqnarray}
\label{H_p}
\fl\qquad\qquad
&\frac{H_p^2(x)}{x^2}  &=
\int_{0}^{H_p(x)} 
\ln y_p\left(\rme^{-\epsilon}\right)\,
\rmd\epsilon - H_p(x) \ln y_p\left(\rme^{-H_p(x)}\right),\\
\label{g_p}
\fl\qquad\qquad
\mbox{and}\quad 
&g_p(x) &= 2\frac{H_p(x)}{x} + 
x \ln y_p\left(\rme^{-H_p(x)}\right),
\end{eqnarray}
respectively.  Thus, given $x$, one has to find $H_p(x)$ by implicitly
solving \eref{H_p}, then substitute it back in \eref{g_p} to get $g_p(x)$,
and finally
\begin{equation}
  \rho_p(E,l) \approx 
  \exp\left[\sqrt{E}\,g_p\left(\frac{l}{\sqrt{E}}\right)\right].
  \label{large deviation}
\end{equation}

For large $x$, using \eref{H_p} and \eref{g_p}, it can be shown that
\begin{equation}
  g_p(x) \approx 2 b(p) -\frac{1}{b(p)} \exp\left[-b(p) x\right]
  \qquad\mbox{as}\quad x\rightarrow \infty,
\end{equation}
where $b(p)$ is given in \eref{beta} and \eref{b(p)}.  Thus, from
\eref{large deviation}, $\rho_p(E)=\rho_p(E,l\rightarrow\infty) \sim
\exp[2b(p)\sqrt{E}]$ to leading order for large $E$, which is the
generalization of the Hardy-Ramanujan formula~\cite{HR} for $\rho_0(E)$,
provided by Meinardus~\cite{Meinardus}.  The normalized cumulative
distribution of $l$, i.e., $C_p(l|E)=\rho_p(E,l)/\rho_p(E)$, for large $E$
and $l\gg\sqrt{E}$, is therefore
\begin{equation}
  C_p(l|E)\approx 
  \exp\left[-\frac{\sqrt{E}}{b(p)}\exp\left(-\frac{b(p)}{\sqrt{E}} l\right)
  \right]
  =  F\left(\frac{b(p)}{\sqrt{E}} \Bigl[l-l^*(E)\Bigr]\right),
\label{C_p}
\end{equation}
where the characteristic value of $l$ is $l^*(E) =
[\sqrt{E}/b(p)]\ln(\sqrt{E}/b(p))$, and the scaling function has the Gumbel
form, $F(z)=\exp[-\exp[-z]]$.  The result for the $p=0$ case, i.e., for
$C_0(l|E)$, was first derived Erd\"os and Lehner~\cite{Erdos}. \Eref{C_p}
provides a generalization of their result, which is valid for all $p$.
The probability distribution $P_p(l|E)=C_p(l|E)-C_p(l-1|E)\approx {\partial
  C_p(l|E)}/{\partial l}$, obtained from \eref{C_p},
\begin{equation}
  \fl\quad
  P_p(l|E)\approx
  \frac{b(p)}{\sqrt{E}}\,
  F'\left(\frac{b(p)}{\sqrt{E}} \Bigl[l-l^*(E)\Bigr]\right),
  \quad
  \mbox{where}\quad  F'(z)= \exp[-z-\exp[-z]],
\label{pdfgumbel}
\end{equation}
is highly asymmetric around the peak at $l=l^*(E)$. This limiting
distribution describes the probability of {\it typical} fluctuations of
$\Or(\sqrt{E})$ of the random variable $l$ around the peak $l^*(E)$.

\section{Summary and remarks}
\label{summary}

In summary, we have obtained a generalized grand partition function for the
minimal difference $p$ partition (\MDPns) of a positive integer $E$, where
smallest part is at least $s$ and largest part is at most $l$, in the
scaling limit $\beta \propto E^{-1/2} \rightarrow 0$, in terms of the
scaling variables $\beta l$ and $\beta s$.  The limit $\beta l\rightarrow
\infty$ also provides a simpler derivation of an earlier result~\cite{CMO},
which showed a link between the exclusion statistics and the \MDP problem,
by showing that both problems are described by the same grand partition
function in the limit $\beta \rightarrow 0$.  Using the grand partition
function we have computed the limiting shape of the Young diagram of the
\MDP problem for all $p$, and also provided a simple physical
interpretation of the result.
Although the Young diagram is defined only for integer values of $p$, one
can analytically continue the expression \eref{shape} for the width $W_h$
of the Young diagram to noninteger values of $p$. For $0<p<1$, $W_h$
corresponds to the number of particles each of which has energy at least
$h$, in a system where the particles obeys exclusion statistics.
We have also obtained the asymptotic distribution of the largest part of
the Young diagram and showed that the scaled distribution has a Gumbel form
for all $p$. When one analytically continues, for $0<p<1$, the largest part
corresponds to the highest occupied energy level in exclusion statistics.

Note that for $p=0$, the transposed Young diagram of a given partition
gives another valid $p=0$ partition.  This symmetry implies that the
statistics of the largest part is the same as the statistics of the number
of parts in the $p=0$ partition problem.  The distribution of the number of
parts for $p=0$ was computed by Erd\"os and Lehner~\cite{Erdos} and in the
appropriate scaling limit it has a Gumbel form. However, the symmetry
between the number of parts and the largest part no longer holds when
$p>0$, where the distribution of the number of parts become Gaussian (see
\cite{CMO} and references therein).

Recently, the statistics of the number of parts for a general partitions of
the form $E=\sum n_{i}i^{1/\nu}$ that corresponds to having a power-law
density of states, $\tilde{\rho}(\epsilon)\sim \epsilon^{\nu-1}$, has been
studied~\cite{CLM} in the bosonic sector ($p=0$).  Clearly, $\nu =1$
corresponds to the usual unrestricted partition problem, where the number
of parts obey Gumbel statistics.  Interestingly, for $\nu\neq 1$, the
authors in~\cite{CLM} also obtained the other two universal distribution
laws of extreme value statistics, namely the Fr\'echet and Weibull
distributions for $0<\nu<1$ and $\nu>1$ respectively.

Therefore, the general partition problem can be defined in the parameter
space of $(\nu,p)$ with $\nu>0$ and $p\ge0$. In this parameter space the
point $(\nu=1,p=0)$ is a very special one at which both the number of parts
and the largest part obey the same statistics given by the Gumbel
distribution.  Along the line $\nu=1$, the limiting distribution of the
number of parts becomes Gaussian as soon as $p>0$, whereas the limiting
distribution of the largest part remains Gumbel for all $p$, as we have
shown in this paper.  On the other hand, along the $p=0$ line, for the
number of parts one finds~\cite{CLM} all the three universal laws of the
extreme value statistics, for the parameter $0<\nu<1$, $\nu=1$, and $\nu>1$
respectively.  Therefore, it is interesting to ask whether there is any
region in the $(\nu,p)$ parameter space, where the largest part obeys
another statistics than the Gumbel one. The answer is negative.  For a
general density of states, \eref{action} includes a factor of
$\tilde{\rho}(\epsilon/\beta)$ in the integrand. Following the similar
steps provided afterwards, it can be shown that even for the power-law
density of states $\tilde{\rho}(\epsilon)\sim \epsilon^{\nu-1}$, the scaled
distribution of the largest part remains Gumbel in the whole $(\nu,p)$
plane.  Thus, the largest part obeys a more robust law, in contrast to the
number of parts.

\addcontentsline{toc}{section}{\protect\numberline{}Note added in proof}
\section*{Note added in proof} 

We thank K. Hikami for pointing out \cite{hikami} in which the author
obtained the solution of a recursion relation similar to \eref{frel1} with
$s=0$ for arbitrary $\beta$. However, for the purpose of this paper we
require the solution only in the limit $\beta\rightarrow 0$. In this limit
it is simpler to obtain it using the method presented in this paper rather
than obtaining by taking the limit $\beta\rightarrow 0$ in the solution of
\cite{hikami}.  The average occupation number at a level $i$ for the
exclusion statistics has been studied in \cite{DO1, DO2, Wu, hikami2,
  hikami3}, which also can be obtained from \eref{number of columns} simply
through
\begin{equation}
\langle n_i
\rangle=-\frac{\partial}{\partial s}\left\langle N_s^l
  (z)\right\rangle\Big|_{s=i} =\frac{\partial}{\partial l}\left\langle
  N_s^l (z)\right\rangle\Big|_{l=i},
\end{equation}
which via elementary algebra yields
\begin{equation}
\langle n_i
\rangle = \left(\frac{1}{y_p(z\rme^{-\beta i}) -1} + p \right)^{-1}
\end{equation}

\addcontentsline{toc}{section}{\protect\numberline{}Acknowledgments}
\ack

AC, SNM and SS acknowledge the support of the Indo-French Centre for the
Promotion of Advanced Research (IFCPAR/CEFIPRA) under Project 3404-2.

\section*{References}
\addcontentsline{toc}{section}{\protect\numberline{}\refname}


\begin{thebibliography}{10}


\bibitem{Haldane} Haldane F D M 1991 ``Fractional statistics'' in arbitrary
  dimensions: A generalization of the Pauli principle {\it Phys.  Rev.
    Lett.} {\bf 67} 937


\bibitem{DO1} Dasni\`eres de Veigy A and Ouvry S 1994 Equation of state of
an anyon gas in a strong magnetic field {\it Phys. Rev. Lett.}  {\bf 72}
600

\bibitem{DO2} Dasni\`eres de Veigy A and Ouvry S 1995 One-dimensional
    statistical mechanics for identical particles: the Calogero and anyon
    cases {\it Mod. Phys. Lett.} B {\bf 9} 271

\bibitem{Wu} Wu Y S 1994 Statistical distribution for generalized ideal gas
of fractional-statistics particles {\it Phys. Rev. Lett.} {\bf 73} 922



\bibitem{MS1} Murthy M V N and Shankar R 1994 Haldane exclusion statistics
and second virial coefficient {\it Phys. Rev. Lett.} {\bf 72} 3629


\bibitem{MS2} Murthy M V N and Shankar R 1994 Thermodynamics of a
  one-dimensional ideal gas with fractional exclusion statistics {\it
  Phys. Rev. Lett.} {\bf 73} 3331


\bibitem{Isakov} Isakov S B 1994 Fractional statistics in one dimension:
modeling by means of $1/x^2$ interaction and statistical mechanics {\it
Int. J. Mod. Phys.} A {\bf 9} 2563

\bibitem{Bergere} Berg\`ere M C 2000 Fractional statistic {\it
J. Math. Phys.} {\bf 41} 7252

\bibitem{Andrews} Andrews G E 1998 {\it The Theory of Partitions}
  (Cambridge University Press, Cambridge)


\bibitem{CMO} Comtet A, Majumdar S N and Ouvry S 2007 Integer partitions
  and exclusion statistics {\it J. Phys. A: Math. Theor.} {\bf 40} 11255

\bibitem{Temperley} Temperley H N Y 1952 Statistical mechanics and the
 partition of numbers: the form of crystal surfaces {\it Proc. Cambridge
 Philos. Soc.}  {\bf 48} 683


\bibitem{Vershik} Vershik A M 1996 Statistical mechanics of combinatorial
partitions and their limit shapes {\it Funct. Anal. Appl.} {\bf 30} 90


\bibitem{Vershik2} Freiman G, Vershik A M and Yakubovich Yu V 2000 A local
limit theorem for random strict partitions {\it Theory Probab. Appl} {\bf
44} 453


\bibitem{Vershik3} Vershik A M and Yakubovich Yu V 2001 The limit shape and
  fluctuations of random partitions of naturals with fixed number of
  summands {\it Moscow Math. J.} {\bf 1} 457 


\bibitem{Romik} Romik D 2003 Identities arising from limit shapes of
constrained random partitions {\it Preprint}

\bibitem{Erdos} Erd\"os P and Lehner J 1951 The distribution of the number
of summands in the partitions of a positive integer {\it Duke Math. J.}
{\bf 8} 335


\bibitem{HR} Hardy G H and Ramanujan S 1918 Asymptotic formula{\ae} in
    combinatory analysis {\it Proc. London. Math. Soc.}  {\bf 17} 75

\bibitem{Meinardus} Meinardus G 1954 \"{U}ber partitionen mit
differenzenbedingungen {\it Math. Zeitschr.} {\bf 61} 289



\bibitem{CLM} Comtet A, Leboeuf P and Majumdar S N 2007 Level density of a
  Bose gas and extreme value statistics {\it Phys. Rev. Lett.}  {\bf 98}
  070404

\bibitem{hikami} Hikami K 1995 Character and TBA for an ideal $g$-on gas
{\it Phys. Lett. A} {\bf 205} 364


\bibitem{hikami2} Hikami K 1998 Statistical mechanical interpretation of
  the inverse scattering method: level dynamics for exclusion statistics
  {\it Phys. Rev. Lett.} {\bf 80} 4374


\bibitem{hikami3} Hikami K 2000 Exclusion statistics and chiral partition
  function {\it Physics and Combinatorics (Proc.  Nagoya 2000 Int.
    Workshop)} ed A N Kirillov and N Liskova (Singapore: World Scientific)
  pp~22--48

\end{thebibliography}
\end{document}